\documentclass[conference]{IEEEtran}
\IEEEoverridecommandlockouts
\usepackage{tikz}
\usetikzlibrary{arrows.meta,positioning,calc}

\IEEEtriggeratref{10}
\usepackage{cite}
\usepackage{amsmath,amssymb,amsfonts}
\usepackage{url}
\usepackage[colorlinks=true,linkcolor=black,citecolor=black,urlcolor=blue,hypertexnames=false]{hyperref}
\usepackage{booktabs}
\usepackage{graphicx}
\usepackage{dblfloatfix}
\usepackage{placeins}
\usepackage{float}
\newfloat{algorithm}{tbp}{loa}
\floatname{algorithm}{Algorithm}
\floatstyle{ruled}
\restylefloat{algorithm}

\usepackage{textcomp}
\usepackage{xcolor}
\def\BibTeX{{\rm B\kern-.05em{\sc i\kern-.025em b}\kern-.08em
    T\kern-.1667em\lower.7ex\hbox{E}\kern-.125emX}}
\begin{document}

\title{Joint Network-and-Server Congestion in Multi-Source Traffic Allocation: A Convex Formulation and Price-Based Decentralization\\[-0.15em]
{\large (Extended Version)}}

\author{\IEEEauthorblockN{Tamoghna Sarkar and Bhaskar Krishnamachari}
\IEEEauthorblockA{\textit{Ming Hsieh Department of Electrical and Computer Engineering} \\
\textit{Viterbi School of Engineering, University of Southern California}\\
Los Angeles, CA, USA \\
\{tsarkar, bkrishna\}@usc.edu}}

\maketitle

\begingroup
\renewcommand\thefootnote{}
\footnotetext{This manuscript is an extended and corrected version of the paper published at WiOpt 2026~\cite{wiopt2026jnsc}. Appendix~\ref{app:corrections} summarizes the corrections and additional material.}
\addtocounter{footnote}{-1}
\endgroup

\begin{abstract}
This paper studies a rate allocation problem that arises in many networked and distributed systems: steady-state traffic rate allocation from multiple sources to multiple service nodes when both (i) the access-path delay on each source-node route is rate-dependent (capacity-constrained) and convex, and (ii) each service node (also capacity-constrained) experiences a load-dependent queueing delay driven by aggregate load from all sources. We show that the resulting flow-weighted end-to-end delay minimization is a convex program, yielding a global system-optimal solution characterized by KKT conditions that equalize total marginal costs (a path marginal access term plus a node congestion price) across all utilized routes. This condition admits a Wardrop-type interpretation: for each source, all utilized options equalize total marginal cost, while any option with strictly larger total marginal cost receives no flow. Building on this structure, we develop a lightweight distributed pricing-based algorithm in which each service node locally computes and broadcasts a scalar congestion price from its observed aggregate load, while each source updates its traffic split by solving a small separable convex allocation problem under the advertised prices. To preserve shared service-node feasibility under simultaneous source updates, the synchronous implementation coordinates their common update step. We prove that every feasible fixed point of the distributed algorithm is globally system-optimal. Numerical illustrations demonstrate convergence of the distributed iteration to the centralized optimum on a verified static instance, with conservation, capacity, price-consistency, fixed-point, and KKT residuals at numerical precision, and highlight the trade-offs induced by jointly modeling access and service congestion.
\end{abstract}


\section{Introduction}

Many networked systems route traffic from multiple ingress points to one of several service nodes (often with splittable flow), including edge-to-cloud offloading, pub-sub overlays with distributed brokers, and wireless/RAN traffic steering across base stations or compute pools.
A recurring difficulty is that end-to-end latency is shaped by \emph{two coupled congestion effects}: (i) the access path from a source to a chosen node can become congested as more rate is routed on that path, and (ii) the service node itself incurs queueing/processing delay that grows with aggregate load.
Steering traffic to a nearby node may reduce baseline transport delay but overload either the access path or the node; steering to a lightly-loaded node may reduce server delay but increase access delay.
These access--service trade-offs motivate allocation mechanisms that are globally optimal under clear conditions yet implementable with lightweight distributed signaling.

\textbf{Model overview:}
We study a steady-state allocation problem with sources $i\in\mathcal{I}$ of fixed offered rates $\lambda_i$ that split traffic across service nodes $j\in\mathcal{J}$ via $\lambda_{ij}\ge 0$.
Traffic on $(i,j)$ experiences an increasing convex \emph{rate-dependent access delay} $D_{ij}(\lambda_{ij})$, while service node $j$ experiences an increasing convex \emph{load-dependent delay} $D_j(\Lambda_j)$ under aggregate load $\Lambda_j=\sum_i \lambda_{ij}$.
Our objective minimizes total flow-weighted end-to-end delay (formalized in Sec.~\ref{sec:model}), yielding a convex program under standard assumptions.
This convexity enables a transparent optimality characterization in terms of \emph{total marginal cost} equalization.

\section{Related Work}

The problem of allocating traffic across heterogeneous resources sits at the intersection of network utility maximization, distributed load balancing, and modern radio access network (RAN) control. We categorize related literature into three primary domains.

\subsection{Theoretical Foundations and Pricing}
The foundational framework for distributed delay minimization was established by Gallager, who derived necessary conditions for optimal routing based on marginal delay equalization \cite{gallager1977minimum}. This gradient-based approach was later generalized into the Network Utility Maximization (NUM) framework by Kelly \textit{et al.} \cite{kelly1998rate} and Low \textit{et al.} \cite{low1999optimization}, which interprets congestion signals as shadow prices to coordinate source rates. 

While these works primarily focus on transport capacity, recent economic interpretations have bridged the gap between user equilibrium and system optimality. Rambha \textit{et al.}~\cite{rambha2018marginal} demonstrated that marginal cost pricing is the necessary mechanism to align selfish routing with global system optima in transportation networks. From a game-theoretic perspective, Paccagnan \textit{et al.}~\cite{paccagnan2019nash} characterize the relationship between Nash and Wardrop equilibria in aggregative games with coupling constraints and propose decentralized algorithms that converge to these equilibrium notions, providing useful context for Wardrop-type interpretations under shared constraints. Our work extends these ideas by introducing a joint cost model that couples rate-dependent access delays with load-dependent server queueing delays and yields a Wardrop-type total marginal-cost equalization condition on a bipartite source-server topology.

Complementary to our model-based approach, Vu \textit{et al.}~\cite{vu2021fast} study fast routing under uncertainty via adaptive learning in (nonatomic) congestion games with exponential weights, offering a learning-theoretic viewpoint on routing dynamics when costs fluctuate or are only observed implicitly.

\subsection{Wireless and O-RAN Traffic Steering}
In the context of 5G/6G and Open RAN (O-RAN), traffic steering is often treated as a complex resource management problem. Nguyen \textit{et al.} proposed a multi-layer optimization framework for O-RAN that accounts for interference and functional splits, though the resulting non-convexity requires computationally intensive solutions \cite{nguyen2024network}. Conversely, data-driven approaches, such as the Deep Reinforcement Learning (DRL) steering mechanisms proposed by Habib \textit{et al.} \cite{habib2023traffic}, offer flexibility against unknown network dynamics but lack interpretability and provable convergence guarantees. 

Our approach offers a middle ground: a convex ``white-box'' formulation with a scalar congestion price per server. The synchronous distributed implementation preserves shared-capacity feasibility by coordinating aggregate proposed-load directions through a common step.

\subsection{Edge Computing and Load Balancing}
In distributed computing, load balancing strategies have evolved from queue-length-based heuristics, such as the ``Power of Two Choices'' \cite{mitzenmacher2001power} and Join-Idle-Queue (JIQ) \cite{lu2011jiq}, to more rigorous optimization methods. While heuristics are asymptotically optimal in large-scale homogeneous systems, they often struggle with the heterogeneity inherent in edge-cloud continuums.

In large operational networks, PLB~\cite{qureshi2022plb} demonstrates that simple congestion signals can be highly effective for load balancing by triggering path changes in response to congestion, supporting the broader premise that lightweight scalar signals can coordinate distributed traffic allocation. Recent work by Balseiro \textit{et al.}~\cite{balseiro2025load} addresses heterogeneity by applying distributed gradient descent to load balancing with network latencies, a model closely related to ours, though their focus lies heavily on stability analysis under delayed feedback. Similarly, Urgaonkar \textit{et al.}~\cite{urgaonkar2015dynamic} employed Lyapunov optimization for dynamic workload scheduling, and Nguyen \textit{et al.}~\cite{nguyen2021price} explored market-equilibrium approaches for fair resource sharing among competing slices. In contrast, our work prioritizes the minimization of total flow-weighted system delay through a unified marginal cost structure, providing a centralized benchmark and a distributed fixed-point reference whose feasible fixed points are globally optimal.

\subsection{Motivating applications}
Although the formulation in this paper is intentionally abstract, it captures a common set of trade-offs across many practical systems.
We highlight four motivating examples to illustrate the generality of this formulation, including wireless-network scenarios.

\paragraph{Traffic steering and load balancing across small cells in dense wireless HetNets}
In dense 5G and 6G deployments, traffic aggregates, or UEs and gateways, can be steered toward different candidate base stations, such as macro cells and small cells, or toward different sectors.
Here, $D_{ij}(\lambda_{ij})$ can represent an access-side delay that grows with the offered load on the chosen radio and backhaul route, while $D_j(\Lambda_j)$ captures scheduling or queueing delay that rises as a cell or compute pool approaches saturation.

\paragraph{Cloud-RAN and vRAN processing allocation across DU and CU pools}
In virtualized RAN architectures, radio units (sources) may send baseband workloads to one of several DU and CU pools or accelerator clusters (servers) over fronthaul and midhaul.
The function $D_{ij}(\lambda_{ij})$ captures load-dependent transport or buffering delay on the corresponding fronthaul/midhaul path, while $D_j(\Lambda_j)$ captures compute queueing or processing delay at pool $j$ driven by aggregate workload.

\paragraph{Pub-Sub traffic allocation across geographically distributed brokers}
Publishers (sources) such as IoT devices may route messages to one of several brokers (servers) that differ in network distance and in access-path conditions.
The function $D_{ij}(\lambda_{ij})$ captures publisher-to-broker access delay that increases as more traffic is routed along that path (e.g., due to ingress bottlenecks or intermediate queues), while $D_j(\Lambda_j)$ captures broker queueing or processing delay as aggregate publish rate $\Lambda_j$ grows.

\paragraph{Edge-to-cloud offloading across multiple compute sites}
Edge nodes (sources) offload tasks to a set of cloud or edge data centers (servers).
The function $D_{ij}(\lambda_{ij})$ captures uplink and transport delay that can increase with the offered rate on the selected access route, while $D_j(\Lambda_j)$ captures compute queueing delay at site $j$ driven by aggregate offloaded workload.

These examples share a common optimization structure and tradeoff: Traffic should be directed to favorable access paths when possible, but must also be distributed to prevent both access bottlenecks and overloaded service nodes from dominating end-to-end latency. The formulation was originally motivated by a practical pub-sub traffic-allocation problem encountered in an industrial setting. The same optimization structure also arises in the wireless and distributed-computing applications described above.

\subsection{Results and contributions}
This paper provides both centralized and distributed methods for solving~\eqref{eq:objective}.
First, we characterize the globally optimal allocation via convex optimization and KKT conditions, which yields a clear \emph{total marginal cost} structure.
Servers used by a given source equalize the sum of a path-dependent marginal access term and a server-dependent marginal congestion price.
This structural view leads to a pricing interpretation: Each service node computes and broadcasts a scalar congestion price as a function of its current aggregate load.
Second, we develop a lightweight distributed algorithm based on this interpretation.
Service nodes iteratively update and broadcast their congestion prices, and sources update their traffic splits by solving small separable convex allocation problems under the advertised prices.
The distributed implementation uses one congestion price per service node and a common step computed from aggregate proposed-load directions to preserve shared-capacity feasibility.
Every feasible fixed point of the distributed algorithm is globally optimal, and we evaluate its convergence empirically.

While the algorithmic mechanics leverage established principles of network utility maximization and pricing-based distributed iteration, the distinct contribution of this work is the structural synthesis of rate-dependent access delays and load-dependent service delays into a unified framework. This formulation yields three specific benefits over generic optimization approaches: 1) it captures the specific architectural constraints of modern edge, O-RAN and cloud distributed systems where access and compute bottlenecks are coupled; 2) it exposes an interpretable scalar-price interface together with the additional coordination needed to guarantee shared-capacity feasibility; and 3) it provides a globally optimal centralized benchmark and a fixed-point optimality certificate against which domain-specific heuristic load-balancing strategies can be rigorously evaluated.

\subsection{Paper organization}
The remainder of the paper is organized as follows.
Section~\ref{sec:model} describes the system model and formalizes the traffic allocation problem.
Section~\ref{sec:centralized} analyzes the convex program, derives the optimality conditions, and presents the centralized solution characterization.
Section~\ref{sec:distributed} introduces the distributed price-based algorithm and establishes the optimality of its feasible fixed points.
Section~\ref{sec:results} provides numerical illustrations demonstrating the effectiveness of the proposed approach under representative parameter settings.
Finally, Section~\ref{sec:conclusion} concludes and outlines directions for future work.

\begin{figure*}[t]
\centering

\begin{tikzpicture}[
  font=\small,
  >=Latex,
  src/.style={draw, rounded corners, thick, minimum width=18mm, minimum height=9mm, align=center},
  netq/.style={draw, circle, thick, minimum size=7mm, inner sep=0pt},
  srv/.style={draw, rounded corners, thick, minimum width=18mm, minimum height=9mm, align=center},
  edgelab/.style={midway, sloped, fill=white, inner sep=1pt, font=\scriptsize}
]

\coordinate (Sx) at (0,0);
\coordinate (Qx) at (4,0);
\coordinate (Jx) at (8,0);




\node[anchor=east] at (3.5,  2) {$\lambda_1$};
\node[anchor=east] at (3.5,  0) {$\lambda_2$};
\node[anchor=east] at (3.5, -2) {$\lambda_3$};

\node[netq] (q1) at (4,  2) {src 1};
\node[netq] (q2) at (4,  0) {src 2};
\node[netq] (q3) at (4, -2) {src 3};

\node[srv] (j1) at (8,  2) {Server 1};
\node[srv] (j2) at (8,  0) {Server 2};
\node[srv] (j3) at (8, -2) {Server 3};


\node[anchor=west, font=\footnotesize, align=left] at (-2,1.5)
{\textbf{At Sources:}\\ Flow splitting: $\sum_{j} \lambda_{ij} = \lambda_i$\\
Access delay per path: $d_{ij} = D_{ij}(\lambda_{ij})$\\
Maximum capacity per path $\lambda_{ij} < \mu_{ij}$};

\draw[->, thick] (q1) -- node[edgelab, above] {$\lambda_{1,1}$} (j1.west);
\draw[->, thick] (q1) -- node[edgelab, above] {$\lambda_{1,2}$} (j2.west);
\draw[->, thick] (q1) -- node[edgelab, below] {$\lambda_{1,3}$} (j3.west);

\draw[->, thick] (q2) -- node[edgelab, above] {$\lambda_{2,1}$} (j1.west);
\draw[->, thick] (q2) -- node[edgelab, above] {$\lambda_{2,2}$} (j2.west);
\draw[->, thick] (q2) -- node[edgelab, below] {$\lambda_{2,3}$} (j3.west);

\draw[->, thick] (q3) -- node[edgelab, above] {$\lambda_{3,1}$} (j1.west);
\draw[->, thick] (q3) -- node[edgelab, above] {$\lambda_{3,2}$} (j2.west);
\draw[->, thick] (q3) -- node[edgelab, below] {$\lambda_{3,3}$} (j3.west);


\node[anchor=west, font=\footnotesize, align=left] at (9.1,1.5)
{\textbf{At Servers:}\\Flow Aggregation:  $\Lambda_j=\sum_i \lambda_{ij}$\\
Server delay: $D_j(\Lambda_j)$\\
Maximum server capacity $\Lambda_{j} < \mu_{j}$};




\end{tikzpicture}
\caption{Illustration of the general formulation (paper notation). Each source $i$ with rate $\lambda_i$ splits traffic into $\lambda_{ij}$ over servers $j$. Each path has capacity $\mu_{ij}$ and experiences a convex increasing access delay $D_{ij}(\lambda_{ij})$. Each server $j$ experiences load $\Lambda_j=\sum_i \lambda_{ij}$ and a convex increasing delay $D_j(\Lambda_j)$, with capacity $\mu_j$.}
\label{fig:general_problem_tikz}

\end{figure*}

\section{System Model and Problem Formulation}
\label{sec:model}

We consider a system with multiple independent traffic sources and multiple service nodes (SN).
The model represents a steady-state load allocation problem in which traffic from multiple sources is routed to heterogeneous service nodes with finite capacities.
The delay here consists of a path-dependent queueing component that depends on the traffic placed on each source to service-node path and the traffic arriving to a service-node queueing component that depends on aggregate load at each SN.

\subsection{System Model}
\label{subsec:system_model}

Let $\mathcal{I} = \{1,\dots,m\}$ denote the set of traffic sources and $\mathcal{J} = \{1,\dots,n\}$ the set of service nodes.
Each source $i \in \mathcal{I}$ generates traffic at a fixed rate $\lambda_i > 0$.
Traffic from a source may be split arbitrarily across service nodes.
Let $\lambda_{ij} \ge 0$ denote the rate at which source $i$ sends traffic to service node $j$.

The routing variables $\{\lambda_{ij}\}$ are the decision variables of the system and satisfy the flow conservation constraints for each source $i$: $\sum_{j \in \mathcal{J}} \lambda_{ij} = \lambda_i$. 
Each source-node path $(i,j)$ is associated with a finite access capacity $\mu_{ij} > 0$.
We restrict attention to feasible operating points satisfying $\lambda_{ij} < \mu_{ij}$ for all $(i,j)$.
This captures access bottlenecks such as last-mile links, ingress queues, or admission limits that become unstable as the routed rate approaches $\mu_{ij}$.

Each service node $j \in \mathcal{J}$ provides service at rate $\mu_j > 0$.
The aggregate arrival rate at node $j$ is specified as: $\Lambda_j \triangleq \sum_{i \in \mathcal{I}} \lambda_{ij}$,
which must satisfy a stability condition $\Lambda_j < \mu_j$.

\subsection{Delay Model}
\label{subsec:delay_model}

Traffic routed from source $i$ to service node $j$ experiences two queueing components.

\paragraph{Path-dependent access delay}
Let $D_{ij}(\cdot)$ denote the access (or path) delay function on the route from source $i$ to service node $j$, modeled as a function of the routed rate $\lambda_{ij}$.
We assume $D_{ij}(\lambda_{ij})$ is continuous, twice-differentiable, strictly increasing, and convex over the domain $\lambda_{ij} < \mu_{ij}$, and satisfies $\lim_{\lambda_{ij} \uparrow \mu_{ij}} D_{ij}(\lambda_{ij}) = +\infty$.

A canonical example is the M/M/1 access-delay model,
\begin{equation}
    D_{ij}(\lambda_{ij}) = \frac{1}{\mu_{ij} - \lambda_{ij}}.
    \label{eq:mm1_access}
\end{equation}
Our analysis applies to any access delay function satisfying the above properties.

\paragraph{SN queueing delay}
Let $D_j(\cdot)$ denote the queueing delay function at service node $j$, which depends on the aggregate arrival rate. We model the queueing delay at service node $j$ as a function $D_j(\Lambda_j)$ that is continuous, twice-differentiable, strictly increasing, and convex over the domain $\Lambda_j < \mu_j$, and assume $\lim_{\Lambda_j\uparrow\mu_j}D_j(\Lambda_j)=+\infty$.
A canonical example is the M/M/1 model
\begin{equation}
    D_j(\Lambda_j) = \frac{1}{\mu_j - \Lambda_j}.
    \label{eq:mm1_node}
\end{equation}
though our analysis applies to any delay function satisfying the above properties.

\paragraph{End-to-end delay}
The total end-to-end delay experienced by traffic from source $i$ to service node $j$ is
\begin{equation}
    \widetilde{D}_{ij}(\lambda) = D_{ij}(\lambda_{ij}) + D_j(\Lambda_j).
    \label{eq:e2e_delay}
\end{equation}

\subsection{Optimization Objective}
\label{subsec:objective}

Given the system model and delay structure described above, traffic routed from source $i$ to service node $j$ experiences an end-to-end delay $\widetilde{D}_{ij}(\lambda)$.
The system-wide performance depends on how traffic is split across service nodes and on the access and SN queueing delays induced by the resulting routed and aggregate loads.
We therefore consider an objective that minimizes the total delay experienced by all traffic in the system, weighted by the corresponding flow rates.

Formally, we consider the following optimization problem:
\begin{align}
    \min_{\{\lambda_{ij}\}} \quad
    & \sum_{i \in \mathcal{I}} \sum_{j \in \mathcal{J}} \lambda_{ij} \, \widetilde{D}_{ij}(\lambda)
    \label{eq:objective} \\
    \text{s.t.} \quad
    & \sum_{j \in \mathcal{J}} \lambda_{ij} = \lambda_i, \quad \forall i \in \mathcal{I}, \\
    & \lambda_{ij} \ge 0, \quad \forall i \in \mathcal{I}, \; j \in \mathcal{J}, \\
    & \sum_{i \in \mathcal{I}} \lambda_{ij} < \mu_j, \quad \forall j \in \mathcal{J}, \\
    & \lambda_{ij} < \mu_{ij}, \quad \forall i \in \mathcal{I}, \; j \in \mathcal{J}.
\end{align}

This objective captures the aggregate delay experienced by all traffic in the system and defines a system-optimal routing policy.
We assume that the resulting open stability domain is nonempty. The strict capacity inequalities are part of the effective domain rather than closed constraints with KKT multipliers; a small margin $\epsilon$ is used only by the numerical solvers.
Under the stated assumptions on the delay functions $D_{ij}(\cdot)$ and $D_j(\cdot)$, the problem is convex in the routing variables over the feasible region, as established in the next section.

\section{Centralized System-Optimal Solution}
\label{sec:centralized}

We next characterize the structure of the system-optimal solution under centralized control for the formulation in Section~\ref{sec:model}, in which end-to-end delay includes both a path-dependent access delay that depends on $\lambda_{ij}$ and a SN queueing delay that depends on $\Lambda_j$.

\subsection{Convexity and Existence}
\label{subsec:convexity}

We establish convexity of the optimization problem~\eqref{eq:objective} by rewriting the objective as a sum of separable convex terms.
Recalling that $\widetilde{D}_{ij}(\lambda)=D_{ij}(\lambda_{ij})+D_j(\Lambda_j)$ and $\Lambda_j=\sum_{i\in\mathcal I}\lambda_{ij}$, the objective can be expressed as
\begin{align}
F(\lambda)
&=
\sum_{i\in\mathcal I}\sum_{j\in\mathcal J}\lambda_{ij}D_{ij}(\lambda_{ij})
+
\sum_{i\in\mathcal I}\sum_{j\in\mathcal J}\lambda_{ij}D_j(\Lambda_j)
\nonumber\\
&=
\sum_{i\in\mathcal I}\sum_{j\in\mathcal J}\lambda_{ij}D_{ij}(\lambda_{ij})
+
\sum_{j\in\mathcal J}\Lambda_j D_j(\Lambda_j).
\label{eq:objective_decomposition}
\end{align}

For each pair $(i,j)$, define
\[
g_{ij}(x) \triangleq x\,D_{ij}(x), \qquad x < \mu_{ij},
\]
and for each service node $j$, define
\[
g_j(x) \triangleq x\,D_j(x), \qquad x < \mu_j.
\]
Under the assumptions in Section~\ref{sec:model}, for $x\ge0$ it follows that
\[
\begin{aligned}
g_{ij}''(x) &= 2D'_{ij}(x) + xD''_{ij}(x) \ge 0, \\
g_{j}''(x)  &= 2D'_{j}(x)  + xD''_{j}(x)  \ge 0.
\end{aligned}
\]
so $g_{ij}(\cdot)$ is convex on $x<\mu_{ij}$ and $g_j(\cdot)$ is convex on $x<\mu_j$.
Since each $\lambda_{ij}$ enters the first term of~\eqref{eq:objective_decomposition} through the convex function $g_{ij}(\lambda_{ij})$, the access-delay component is convex in $\{\lambda_{ij}\}$.
Moreover, each aggregate arrival rate $\Lambda_j$ is an affine function of $\{\lambda_{ij}\}_{i\in\mathcal I}$, so the composition $g_j(\Lambda_j)=\Lambda_jD_j(\Lambda_j)$ is convex in the routing variables.
Summing convex terms preserves convexity, implying that the objective in~\eqref{eq:objective} is convex in $\{\lambda_{ij}\}$ over the feasible region.

The feasible set defined by the flow conservation, nonnegativity, and stability constraints is convex.
For queueing delay models of interest, including the M/M/1 form in~\eqref{eq:mm1_access} and~\eqref{eq:mm1_node}, the terms $g_{ij}(\lambda_{ij})$ and $g_j(\Lambda_j)$ diverge as $\lambda_{ij}\uparrow \mu_{ij}$ and $\Lambda_j\uparrow \mu_j$, respectively, ensuring that any minimizing sequence remains strictly within the stability region.
Therefore, the optimization problem~\eqref{eq:objective} admits an optimal solution.

Because the capacity boundary is excluded and the objective diverges there, an optimum lies in the capacity interior. The remaining equality and nonnegativity constraints are affine, so the Karush-Kuhn-Tucker (KKT) conditions below are necessary and sufficient for global optimality.
In the next subsection, we use these conditions to characterize the structure of the system-optimal routing policy.

\subsection{Optimality Conditions and Structural Properties}
\label{subsec:kkt}

We now characterize the structure of the system-optimal solution by deriving the KKT conditions.
These conditions yield an interpretation of optimal routing in terms of \emph{total marginal delay costs} that include both access and SN congestion effects.

\subsubsection*{Lagrangian and KKT Conditions}

Let $\alpha_i$ denote the Lagrange multiplier associated with the flow conservation constraint
$\sum_{j\in\mathcal J}\lambda_{ij}=\lambda_i$ for source $i$, and let $\beta_{ij}\ge 0$ denote the multiplier associated with the non-negativity constraint $\lambda_{ij}\ge 0$.
(We treat the strict stability constraints $\lambda_{ij}<\mu_{ij}$ and $\Lambda_j<\mu_j$ as part of the domain; for the queueing models of interest the objective diverges at the boundary, so the optimum lies in the interior.)

The Lagrangian is given by
\begin{align}
\mathcal{L}(\lambda,\alpha,\beta)
&=
\sum_{i\in\mathcal I}\sum_{j\in\mathcal J} \lambda_{ij} D_{ij}(\lambda_{ij})
+
\sum_{j\in\mathcal J} \Lambda_j D_j(\Lambda_j)
\nonumber\\
&\quad
+
\sum_{i\in\mathcal I} \alpha_i
\Bigl(\lambda_i - \sum_{j\in\mathcal J} \lambda_{ij}\Bigr)
-
\sum_{i\in\mathcal I}\sum_{j\in\mathcal J} \beta_{ij}\lambda_{ij},
\label{eq:lagrangian}
\end{align}
where $\Lambda_j=\sum_{i\in\mathcal I}\lambda_{ij}$.

Stationarity with respect to $\lambda_{ij}$ yields
\begin{equation}
\label{eq:stationarity}
\begin{aligned}
&\Bigl[D_{ij}(\lambda_{ij})+\lambda_{ij} D'_{ij}(\lambda_{ij})\Bigr]
+\Bigl[D_j(\Lambda_j)+\Lambda_j D'_j(\Lambda_j)\Bigr] \\
&\qquad - \alpha_i - \beta_{ij} = 0, \qquad \forall i,j.
\end{aligned}
\end{equation}

Together with primal feasibility, dual feasibility $\beta_{ij}\ge 0$, and complementary slackness
\[
\beta_{ij}\lambda_{ij}=0,
\]
these conditions fully characterize the system-optimal solution.

\subsubsection*{Total Marginal Delay Equalization}

Define the \emph{marginal access cost} on path $(i,j)$ as
\begin{equation}
\label{eq:marginal_access_cost}
C_{ij}(\lambda_{ij})
\triangleq
D_{ij}(\lambda_{ij})+\lambda_{ij} D_{ij}'(\lambda_{ij}),
\end{equation}
and define the \emph{marginal SN congestion cost} at node $j$ as
\begin{equation}
\label{eq:marginal_cost}
C_j(\Lambda_j)
\triangleq
D_j(\Lambda_j)+\Lambda_j D_j'(\Lambda_j).
\end{equation}
The stationarity condition~\eqref{eq:stationarity} can then be written as
\begin{equation}
C_{ij}(\lambda_{ij})+C_j(\Lambda_j)
\begin{cases}
=\alpha_i, & \text{if }\lambda_{ij}>0,\\
\ge \alpha_i, & \text{if }\lambda_{ij}=0.
\end{cases}
\label{eq:wardrop_structure}
\end{equation}

Equation~\eqref{eq:wardrop_structure} implies that, at optimality, each source $i$ routes traffic only over source-node paths $(i,j)$ that minimize the \emph{sum of marginal access cost and marginal SN congestion cost}.
All service nodes that receive positive traffic from a given source $i$ have equal total marginal cost $C_{ij}(\lambda_{ij})+C_j(\Lambda_j)$, while unused paths incur no smaller total marginal cost.

\textit{Wardrop-type Interpretation:} The condition~\eqref{eq:wardrop_structure} admits a Wardrop-type interpretation in terms of \emph{total marginal costs}.
For a given source, all utilized options equalize total marginal cost, while any option with strictly larger total marginal cost receives no flow.
This equalization arises as a consequence of system-optimal routing for a convex objective, without invoking any equilibrium or selfish-routing assumptions.

We treat this centralized solution as a full-information benchmark, against which decentralized implementation can be compared.
The marginal-cost structure in~\eqref{eq:wardrop_structure} will also serve as the basis for the distributed pricing-based solution developed in the next section.

\label{sec:distributed}
Section~\ref{sec:centralized} shows that the system-optimal allocation can be expressed through \emph{total marginal costs}: a path-dependent marginal access term and a service-node (SN) marginal congestion term. We now leverage this structure to define a distributed fixed-point iteration whose feasible fixed points coincide with the centralized optimum. Each SN computes and broadcasts a scalar congestion \emph{price} from its locally observed aggregate load, and each source updates its traffic split by solving a small convex subproblem using locally known access-delay functions and the advertised prices.

\subsection{Pricing interpretation}
\label{subsec:pricing}

From the centralized KKT conditions in Section~\ref{subsec:kkt}, optimality requires total marginal-cost equalization across the options used by each source; see~\eqref{eq:wardrop_structure}.
This suggests the following decentralization: define the congestion \emph{price} at SN $j$ as the marginal congestion cost~\eqref{eq:marginal_cost}, which depends only on the aggregate load at $j$ and is therefore locally computable. Given advertised prices, the remaining marginal access term is local to each source--path pair, so sources can update their allocations using only $(\lambda_i,\{D_{ij}\}_j)$ and the received price vector.

\subsection{Source routing update}
\label{subsec:routing}
Given prices $p$, each source $i$ computes a best-response routing vector by solving the separable convex subproblem
\begin{equation}
\begin{aligned}
\lambda_i^{\rm br}(p)\in\arg\min_{\{\lambda_{ij}\}_{j\in\mathcal J}}\quad &
\sum_{j\in\mathcal J}\bigl[\lambda_{ij}D_{ij}(\lambda_{ij})+p_j\lambda_{ij}\bigr]\\
\text{s.t.}\quad &\sum_j\lambda_{ij}=\lambda_i,\\
&\lambda_{ij}\ge0,\quad \lambda_{ij}<\mu_{ij}.
\end{aligned}
\label{eq:source_br}
\end{equation}
For the M/M/1 access model, its KKT conditions give
\begin{equation}
\lambda_{ij}^{\rm br}(\alpha_i;p)=
\begin{cases}
0, & \alpha_i\le p_j+1/\mu_{ij},\\[2pt]
\mu_{ij}-\sqrt{\mu_{ij}/(\alpha_i-p_j)}, & \alpha_i>p_j+1/\mu_{ij},
\end{cases}
\label{eq:mm1_br}
\end{equation}
where the unique $\alpha_i$ satisfies $\sum_j\lambda_{ij}^{\rm br}(\alpha_i;p)=\lambda_i$. The multiplier search starts at $\min_j(p_j+1/\mu_{ij})$ and solves conservation directly. The resulting best response therefore satisfies the source-rate constraint without any subsequent normalization, preserving the KKT equalities.

Sources form damped best-response directions to avoid overly aggressive simultaneous reactions to changing prices. Because service-node capacity depends on the aggregate direction across all sources, Algorithm~\ref{alg:pricing} selects one common step from the aggregate proposed loads. This preserves source conservation, nonnegativity, access feasibility, and shared service-node feasibility at every update.

\subsection{Service-node price update}
\label{subsec:prices}
Each SN $j$ measures its aggregate arrival rate and computes the induced instantaneous marginal congestion cost. For stability under simultaneous source updates and noisy load estimates, SNs apply the damped price update and broadcast the resulting scalar prices to all sources. After sources form their proposed directions, each SN aggregates its proposed load change and broadcasts the corresponding bound on the common update step.

\subsection{Distributed Algorithm}
\label{subsec:algorithm}

\begin{algorithm}[t]
\caption{Distributed Pricing-Based Routing with Path-Dependent Access Delays}
\label{alg:pricing}
\small
\textbf{Input:} Rates $\{\lambda_i\}$, access delay functions $\{D_{ij}(\cdot)\}$, service delay functions $\{D_j(\cdot)\}$, stepsizes $\{\eta_t\},\{\gamma_t\}$ with $\eta_t,\gamma_t\in(0,1]$, and capacity margin $\delta_s>0$.\\
\textbf{Input:} Initialize feasible routing $\{\lambda_{ij}^{(0)}\}$ with $\Lambda_j^{(0)}\le\mu_j-\delta_s$ and prices $\{p_j^{(0)}\}$.\\[2pt]
\textbf{For $t=0,1,2,\ldots$:}
\begin{enumerate}
\item \emph{Service-node price update (local):} Each SN computes $\Lambda_j^{(t)}=\sum_i\lambda_{ij}^{(t)}$, $\widehat p_j^{(t)}=D_j(\Lambda_j^{(t)})+\Lambda_j^{(t)}D_j'(\Lambda_j^{(t)})$, and $p_j^{(t+1)}=(1-\gamma_t)p_j^{(t)}+\gamma_t\widehat p_j^{(t)}$, then broadcasts $p_j^{(t+1)}$.
\item \emph{Source routing update (local convex subproblem):} Each source $i$ computes $\lambda_i^{\rm br}(p^{(t+1)})$ by solving
\[
\begin{aligned}
\min_{\{\lambda_{ij}\}_{j\in\mathcal J}}\quad &
\sum_{j\in\mathcal J}\left(\lambda_{ij}D_{ij}(\lambda_{ij})+p_j^{(t+1)}\lambda_{ij}\right)\\
\text{s.t.}\quad &\sum_{j\in\mathcal J}\lambda_{ij}=\lambda_i,\\
&\lambda_{ij}\ge0,\quad \lambda_{ij}<\mu_{ij},
\end{aligned}
\]
and forms $\Delta_{ij}^{(t)}=\lambda_{ij}^{\rm br}(p^{(t+1)})-\lambda_{ij}^{(t)}$.
\item \emph{Common feasibility step:} Each SN aggregates $\Delta_j^{(t)}=\sum_i\Delta_{ij}^{(t)}$ and broadcasts
\[
s_j^{(t)}=\begin{cases}
1, & \Delta_j^{(t)}\le0,\\
\min\!\left\{1,\dfrac{\mu_j-\delta_s-\Lambda_j^{(t)}}{\eta_t\Delta_j^{(t)}}\right\}, & \Delta_j^{(t)}>0.
\end{cases}
\]
\item With $s_t=\min_js_j^{(t)}$, apply the original damped best-response update with effective stepsize $\eta_ts_t$:
\[
\lambda_{ij}^{(t+1)}=(1-\eta_ts_t)\lambda_{ij}^{(t)}
+\eta_ts_t\lambda_{ij}^{\rm br}(p^{(t+1)}).
\]
\item Stop when route and price changes, price consistency, and the fixed-point residual satisfy their prescribed tolerances.
\end{enumerate}
\end{algorithm}

The source update is a convex combination and therefore preserves source conservation, nonnegativity, and access feasibility. If $\Delta_j^{(t)}>0$, the common step gives $\Lambda_j^{(t+1)}\le\mu_j-\delta_s$; otherwise the load at SN $j$ does not increase. Hence every iterate remains feasible by construction.

\subsection{Implementation considerations}
\label{subsec:impl_considerations}
Each SN needs its aggregate arrival-rate estimate to compute its scalar price and its aggregate proposed-load direction to determine the common feasibility step. Each source needs only its own offered rate, its access-delay models (or marginal access costs), and the received price and step signals.

Algorithm~\ref{alg:pricing} provides the synchronous distributed reference used in this paper. We leave an asynchronous extension to future work.

\subsection{Fixed-Point Optimality}
\label{subsec:convergence}

A fixed point of Algorithm~\ref{alg:pricing} is a pair
$(\lambda^\star,p^\star)$ such that the induced loads
$\Lambda_j^\star=\sum_i \lambda_{ij}^\star$ satisfy
\begin{equation}
\label{eq:fixedpoint_price}
p_j^\star = D_j(\Lambda_j^\star)+\Lambda_j^\star D_j'(\Lambda_j^\star),
\qquad \forall j\in\mathcal J,
\end{equation}
and each source $i$ is optimal for its local subproblem under prices $p^\star$.
Equivalently, for each $i\in\mathcal I$ there exists a scalar $\alpha_i^\star$
such that
\begin{equation}
\label{eq:fixedpoint_kkt}
C_{ij}(\lambda_{ij}^\star)+p_j^\star
\begin{cases}
=\alpha_i^\star, & \text{if } \lambda_{ij}^\star>0,\\
\ge \alpha_i^\star, & \text{if } \lambda_{ij}^\star=0,
\end{cases}
\qquad \forall j\in\mathcal J.
\end{equation}

\noindent\textbf{Proposition 1 (Optimality of feasible fixed points).}
If $(\lambda^\star,p^\star)$ is feasible and satisfies~\eqref{eq:fixedpoint_price}--\eqref{eq:fixedpoint_kkt}, then $\lambda^\star$ is a globally system-optimal routing solution of~\eqref{eq:objective}.

\smallskip
\noindent\emph{Proof.}
Equation~\eqref{eq:fixedpoint_price} gives $p_j^\star=C_j(\Lambda_j^\star)$. Substitution into~\eqref{eq:fixedpoint_kkt} recovers the centralized KKT conditions in Section~\ref{subsec:kkt}. Convexity and KKT sufficiency imply global optimality. \hfill$\square$

Proposition~1 gives an exact optimality certificate for the distributed algorithm: whenever its iterates reach a feasible fixed point, the resulting allocation is the global centralized optimum. For the M/M/1 price model, the price sensitivity is
\[
\frac{d}{d\Lambda_j}\frac{\mu_j}{(\mu_j-\Lambda_j)^2}
=\frac{2\mu_j}{(\mu_j-\Lambda_j)^3},
\]
which increases as $\Lambda_j$ approaches $\mu_j$. This behavior motivates the damped price update and the common capacity-feasible step in Algorithm~\ref{alg:pricing}. Section~\ref{sec:results} complements the analytical fixed-point guarantee by demonstrating convergence to the centralized optimum on the studied instance and verifying conservation, feasibility, price consistency, fixed-point, and KKT residuals.

\section{Illustrative Results}
\label{sec:results}

This section evaluates the convex flow-weighted allocation objective and the distributed pricing-based iteration on a static multi-source, multi-server topology with heterogeneous access capacities and service rates. The experiments examine the iteration's convergence behavior, compare its limiting allocation with the centralized optimum, and study its response to noisy arrival and service counts in a windowed stochastic experiment.

\subsection{Experimental Setup}
\label{subsec:results_setup}

\paragraph{Static instance}
We use one routing instance with five sources and three service nodes. Four traffic classes have fixed message rates and mean sizes: $50$ msg/s at $2$ KB, $30$ msg/s at $2$ KB, $15$ msg/s at $2$ MB, and $10$ msg/s at $1$ KB. Each source is assigned a fixed subset of these classes; its offered rate is obtained by summing the associated byte rates and converting to MB/s. The resulting source-rate vector is held fixed for both solvers.

\paragraph{Access and service capacities}
The access capacities $\{\mu_{ij}\}$ and service capacities $\{\mu_j\}$ are fixed heterogeneous values generated with the stored random seed. Together with the M/M/1 delay functions, they determine the open stability domain used by the analytical problem.

\paragraph{Initialization}
Both solvers start from a strictly feasible allocation certified by a transportation linear program that enforces source totals and maximizes the minimum capacity headroom. The numerical margin $10^{-8}$ is a solver safeguard, not part of the analytical problem in~\eqref{eq:objective}.

\paragraph{Solver configurations}
The centralized optimum is computed with SLSQP using analytic gradients, followed by an active-set KKT polish and an independent residual evaluation. The distributed solver follows Algorithm~\ref{alg:pricing}; for the M/M/1 model it uses $p_j=\mu_j/(\mu_j-\Lambda_j)^2$ and the exact response in~\eqref{eq:mm1_br}, with no post-hoc normalization. We use $\eta=0.25$, $\gamma=0.5$, at most $4000$ iterations, and a route/price relative-change tolerance of $10^{-10}$. Termination additionally requires conservation, capacity, price-consistency, fixed-point, and KKT residual checks.

\subsection{Centralized Benchmark}
\label{subsec:results_central}
The solver reaches $F^\star=2.0157$ after $29$ SLSQP iterations before polishing. The resulting service-node utilizations are $(0.158136,0.138642,0.102983)$ for SN1--SN3, respectively, and hence remain strictly below one.

\subsection{Distributed Pricing-Based Iteration}
\label{subsec:results_distributed}
We next run Algorithm~\ref{alg:pricing} on the same static instance. The strengthened stopping test is met after $70$ iterations. The distributed objective also rounds to $F_{\mathrm{dist}}=2.0157$; its absolute difference from the centralized value is $1.76\times10^{-13}$ in the stored double-precision execution.

\subsection{Convergence Behavior}
\label{subsec:results_convergence}
\begin{figure}[!t]
  \centering
  \includegraphics[width=\linewidth]{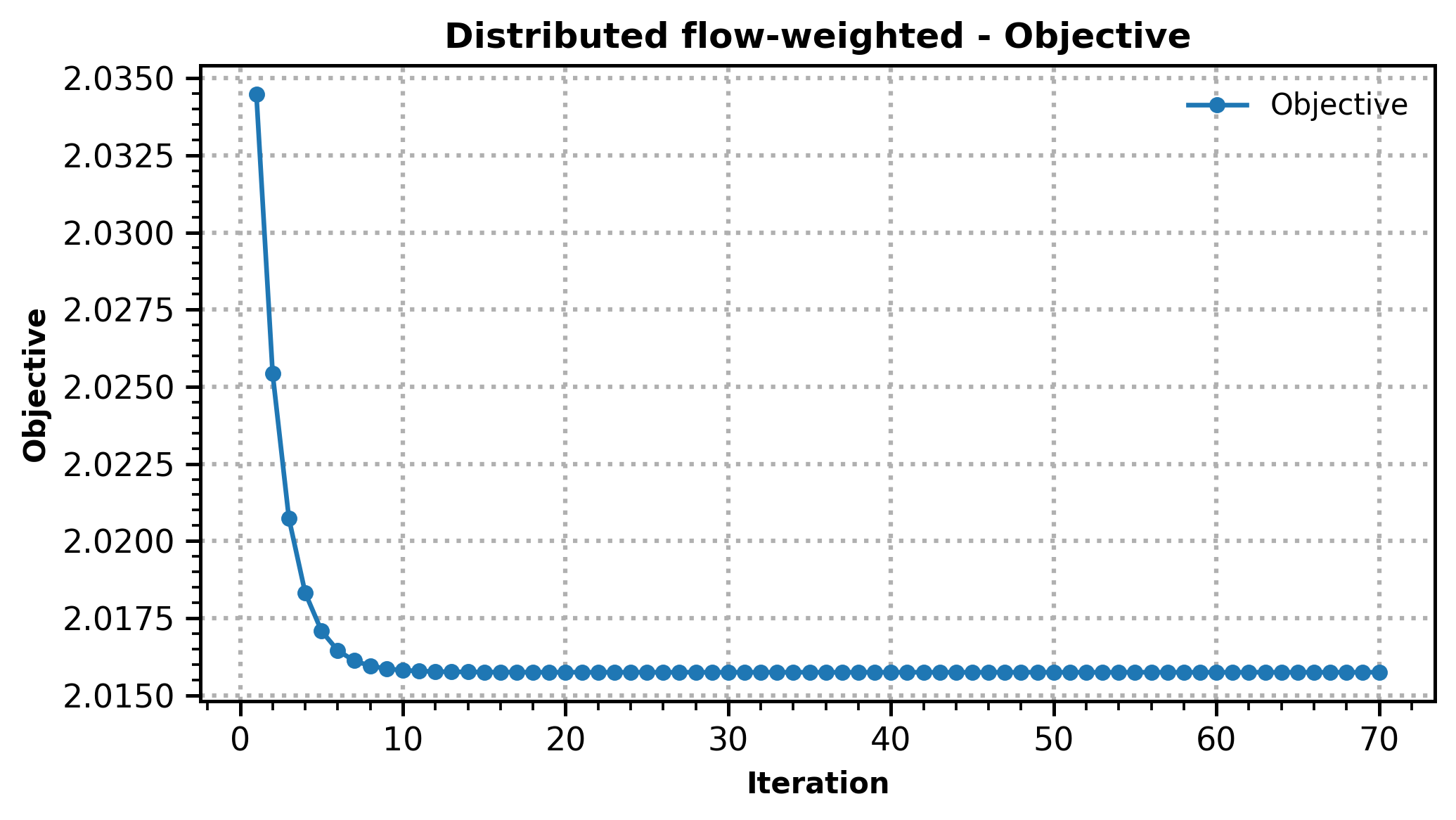}
  \caption{Distributed flow-weighted objective $F(\lambda^{(t)})$ versus iteration. Most of the decrease occurs during the initial transient, after which the curve approaches the centralized value.}
  \label{fig:obj_vs_iter}
\end{figure}

Figure~\ref{fig:obj_vs_iter} plots the objective value $F(\lambda^{(t)})$ across iterations. It decreases from $2.0345$ initially to $2.0158$ within ten iterations and then approaches $2.0157$. Thus most of the improvement occurs during the initial transient, followed by smaller adjustments as the routing and prices settle.

\begin{figure}[!t]
  \centering
  \includegraphics[width=\linewidth]{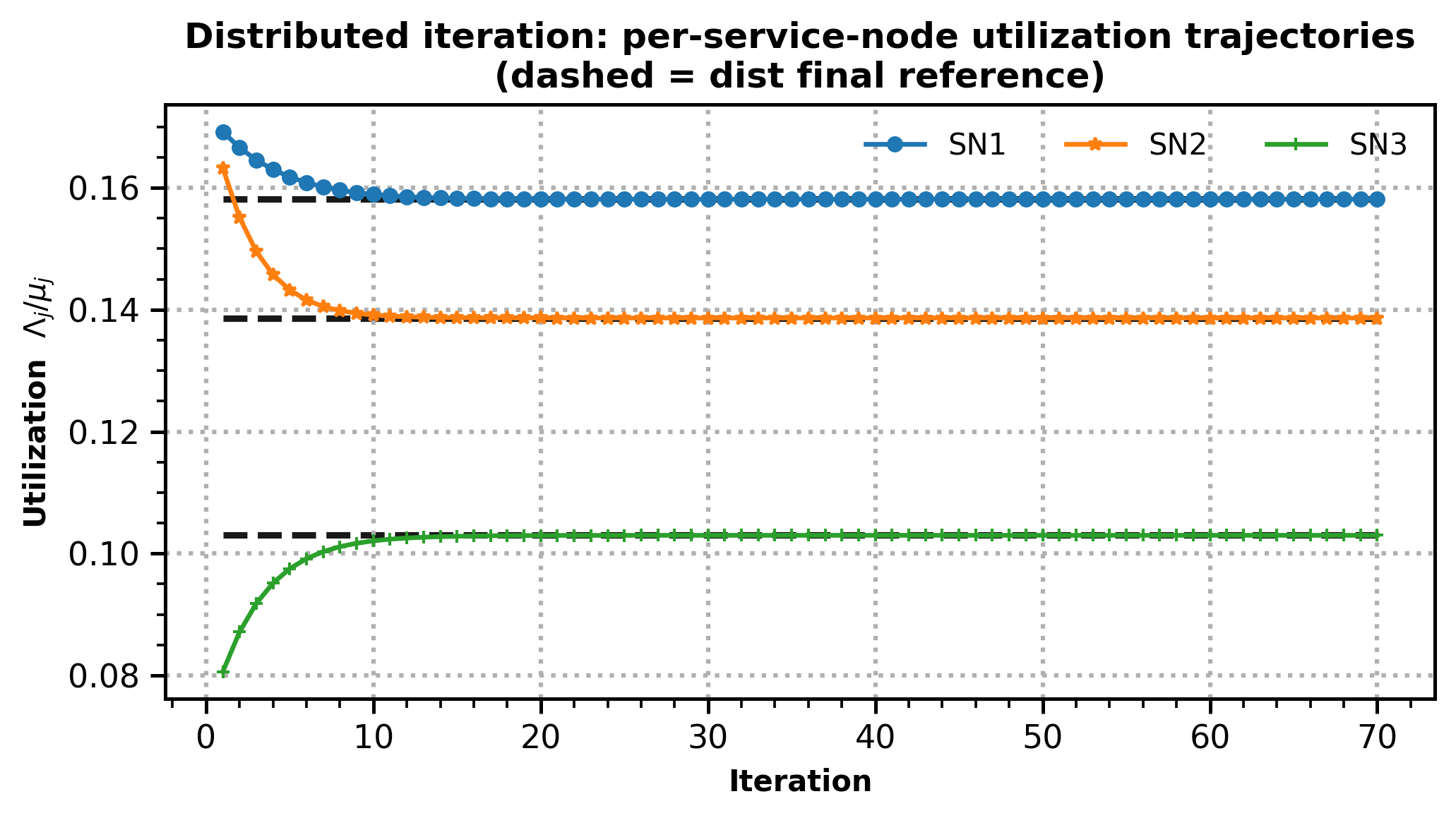}
  \caption{Per-SN utilization trajectories $\Lambda_j^{(t)}/\mu_j$ under the distributed pricing iteration. Dashed lines show the final distributed utilizations.}
  \label{fig:perbroker_utils}
\end{figure}

Figure~\ref{fig:perbroker_utils} reports the per-SN utilization trajectories $\Lambda_j^{(t)}/\mu_j$. SN1 and SN2 move downward from their initial loads, while SN3 absorbs additional traffic; all three stabilize at the final levels marked by the dashed lines. This redistribution illustrates how the scalar congestion prices coordinate simultaneous source updates while maintaining substantial capacity headroom.

\begin{figure}[!t]
  \centering
  \includegraphics[width=\linewidth]{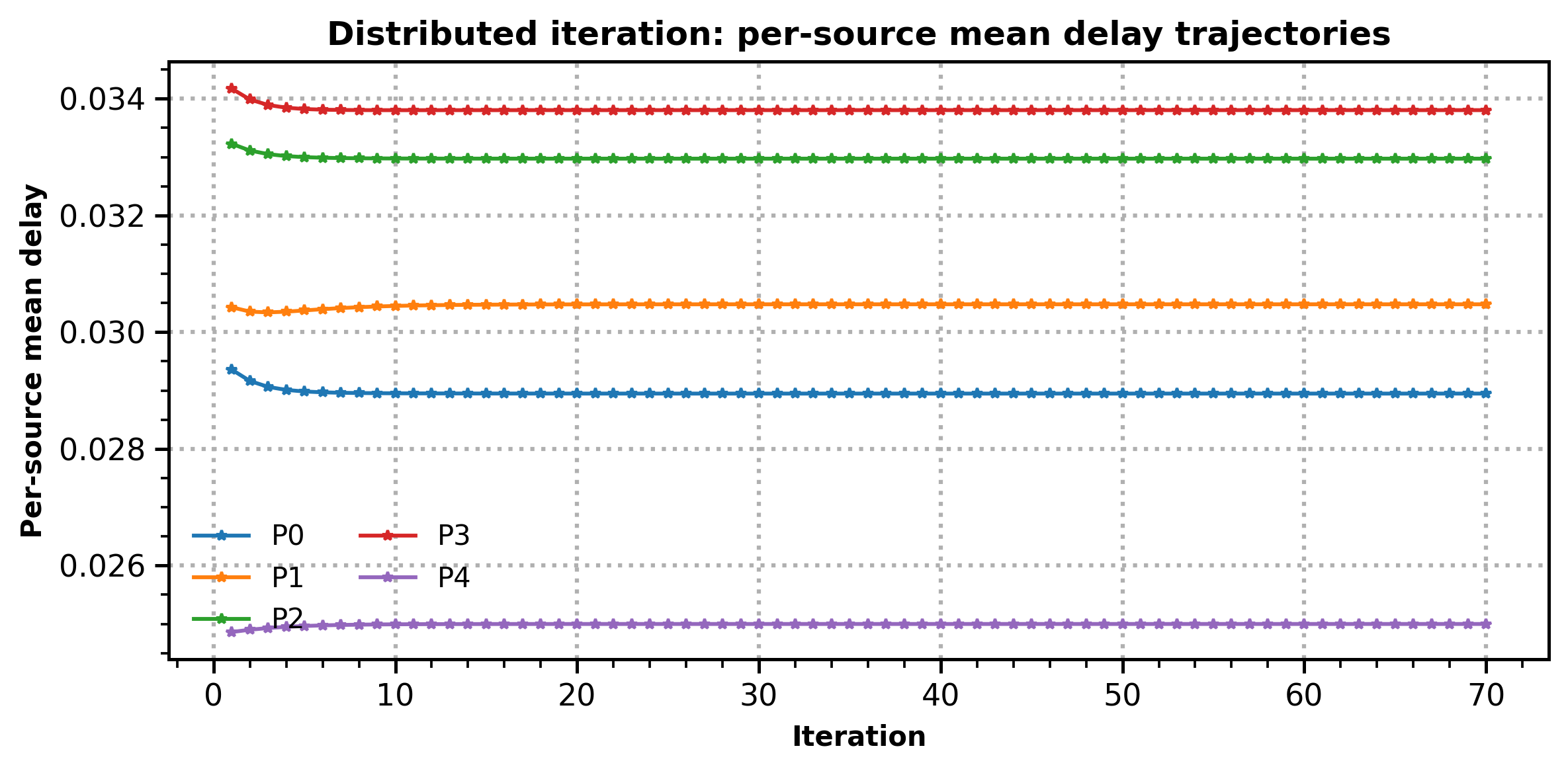}
  \caption{Per-source flow-weighted mean end-to-end delay~\eqref{eq:iter_source_delay} versus iteration. The different plateaus reflect heterogeneous access capacities and routing choices.}
  \label{fig:persource_delay_vs_iter}
\end{figure}

Figure~\ref{fig:persource_delay_vs_iter} plots the per-source flow-weighted mean end-to-end delay
\begin{equation}
\bar D_i^{(t)}=\sum_{j\in\mathcal J}\frac{\lambda_{ij}^{(t)}}{\lambda_i}
\left[D_{ij}(\lambda_{ij}^{(t)})+D_j(\Lambda_j^{(t)})\right].
\label{eq:iter_source_delay}
\end{equation}
The source trajectories exhibit the same short transient and stable plateau as the objective and utilization curves. Their separation reflects heterogeneous access capacities and route choices: P4 attains the smallest final mean delay, while P3 remains the largest. Individual source delays need not all decrease because the algorithm minimizes the system-wide flow-weighted objective rather than each source's delay separately.

\begin{figure*}[!t]
  \centering
  \includegraphics[width=\linewidth]{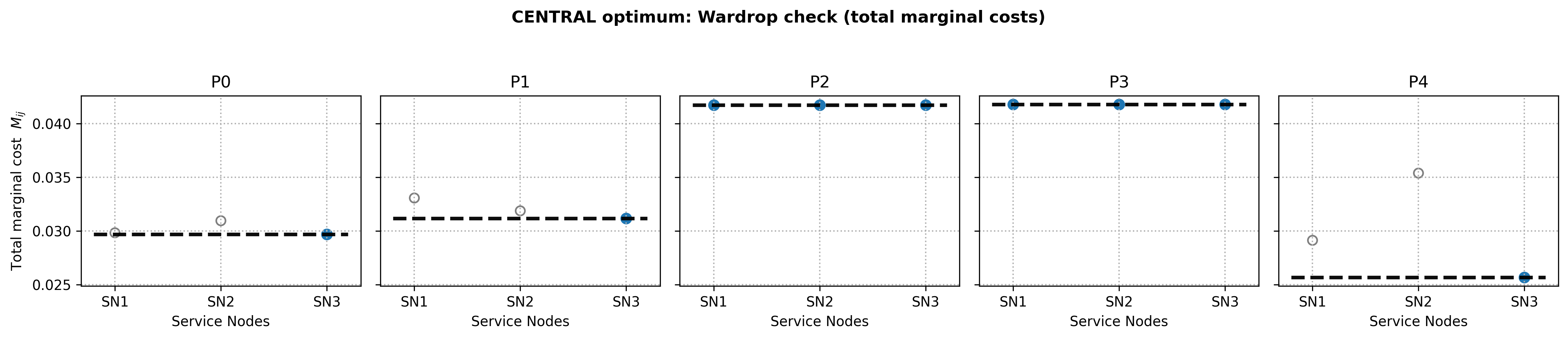}\\[2pt]
  \includegraphics[width=\linewidth]{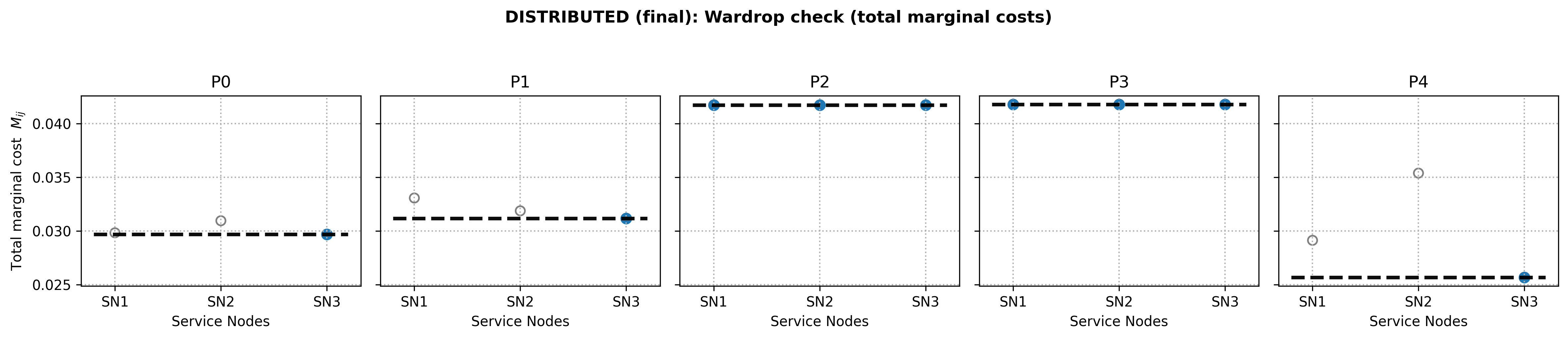}
  \caption{Wardrop/KKT check using total marginal costs~\eqref{eq:numerical_marginal_cost}. For each source, used routes (filled) align closely, while unused routes (hollow) have no smaller marginal cost. Top: centralized optimum. Bottom: distributed fixed point.}
  \label{fig:wardrop_check_central_vs_dist}
\end{figure*}

\subsection{Solution Comparison to Centralized Optimum}
\label{subsec:results_comparison}
The centralized and distributed objectives both round to $2.0157$, and their service-node utilizations both round to $(0.158136,0.138642,0.102983)$. Before rounding, the largest utilization difference between the two allocations is $6.10\times10^{-13}$. The agreement is therefore numerical rather than an assertion that independently computed floating-point arrays are bit-for-bit identical.

Figure~\ref{fig:wardrop_check_central_vs_dist} checks the Wardrop-type structure in~\eqref{eq:wardrop_structure}. For each source and candidate route, it evaluates
\begin{equation}
M_{ij}=C_{ij}(\lambda_{ij})+C_j(\Lambda_j).
\label{eq:numerical_marginal_cost}
\end{equation}
At both solutions, the filled markers corresponding to positive-flow routes align at a common total marginal cost for each source. Hollow markers for unused routes lie at or above that level. In particular, P0, P1, and P4 select SN3, P2 splits across all three service nodes, and P3 uses all three at nearly equal total marginal cost. The centralized and distributed panels reproduce the same active-route pattern.

\begin{table}[!t]
\centering
\scriptsize
\begin{tabular}{lcc}
\toprule
Residual & Central & Distributed\\
\midrule
Conservation & $0$ & $1.31\times10^{-11}$\\
Capacity violation & $0$ & $0$\\
Price consistency & N/A & $1.86\times10^{-14}$\\
Fixed point & $1.23\times10^{-11}$ & $5.86\times10^{-9}$\\
KKT complementarity & $6.94\times10^{-17}$ & $1.41\times10^{-10}$\\
Active stationarity & $6.94\times10^{-18}$ & $5.18\times10^{-12}$\\
\bottomrule
\end{tabular}
\caption{Numerical feasibility and optimality residuals; smaller is better and zero is exact. Price consistency is specific to the distributed price variable and is therefore not applicable to the centralized solve.}
\label{tab:residuals}
\end{table}

Table~\ref{tab:residuals} provides an independent numerical assessment of feasibility and optimality rather than another objective comparison. The conservation residual is the maximum error in a source total, and the capacity-violation residual is the largest positive exceedance of any access-path or service-node capacity; together they test primal feasibility. A reported capacity violation of zero means that no capacity is exceeded, not that the system has no unused capacity.

Price consistency compares the price carried by the distributed iterate with the marginal service cost recomputed from its final aggregate load. It is not applicable to the centralized solver, which has no independent advertised-price variable. The fixed-point residual resolves every source's local best-response problem at the recomputed final prices and measures the resulting allocation discrepancy. Active-route stationarity measures the spread of $M_{ij}$ over routes carrying positive flow from the same source, while KKT complementarity checks that positive flow is assigned only to minimum-marginal-cost routes and that no unused route offers a lower marginal cost.

All entries lie between zero and $5.86\times10^{-9}$ and satisfy the prescribed numerical tolerances. Thus both solutions are feasible, while the distributed endpoint is also consistent with its advertised prices, source best responses, and the centralized KKT optimality conditions. The distributed active-stationarity and KKT-complementarity residuals, $5.18\times10^{-12}$ and $1.41\times10^{-10}$, respectively, agree with the visual equalization in Figure~\ref{fig:wardrop_check_central_vs_dist}. The largest entry is the distributed fixed-point residual, $5.86\times10^{-9}$, which is consistent with terminating an iterative computation rather than polishing the endpoint as a centralized KKT system.

This comparison is stronger than objective agreement alone. Two allocations can have similar objective values without sharing the same route-level optimality structure, particularly near a flat optimum. Here the objective, aggregate utilizations, active-route pattern, marginal costs, and independent residuals all agree. The numerical evidence therefore supports the intended interpretation of the distributed endpoint as the centralized system optimum for this static instance.

\subsection{Windowed Stochastic Count Experiment}
\label{subsec:results_stochastic}
We next test the distributed routing logic under sampled arrival and service counts while retaining the same source rates and capacities. Every five seconds, the experiment draws Poisson arrival counts for each source, routes those counts according to the current splitting probabilities, draws Poisson service-completion budgets for the access paths and service nodes, and updates the aggregate queues. Service nodes form EWMA load estimates and update prices, while sources update their splits after a four-window warmup using the same best-response structure with inertia. This construction introduces finite-window sampling variability into the load measurements and routing updates.

\begin{figure}[!t]
  \centering
  \includegraphics[width=\linewidth]{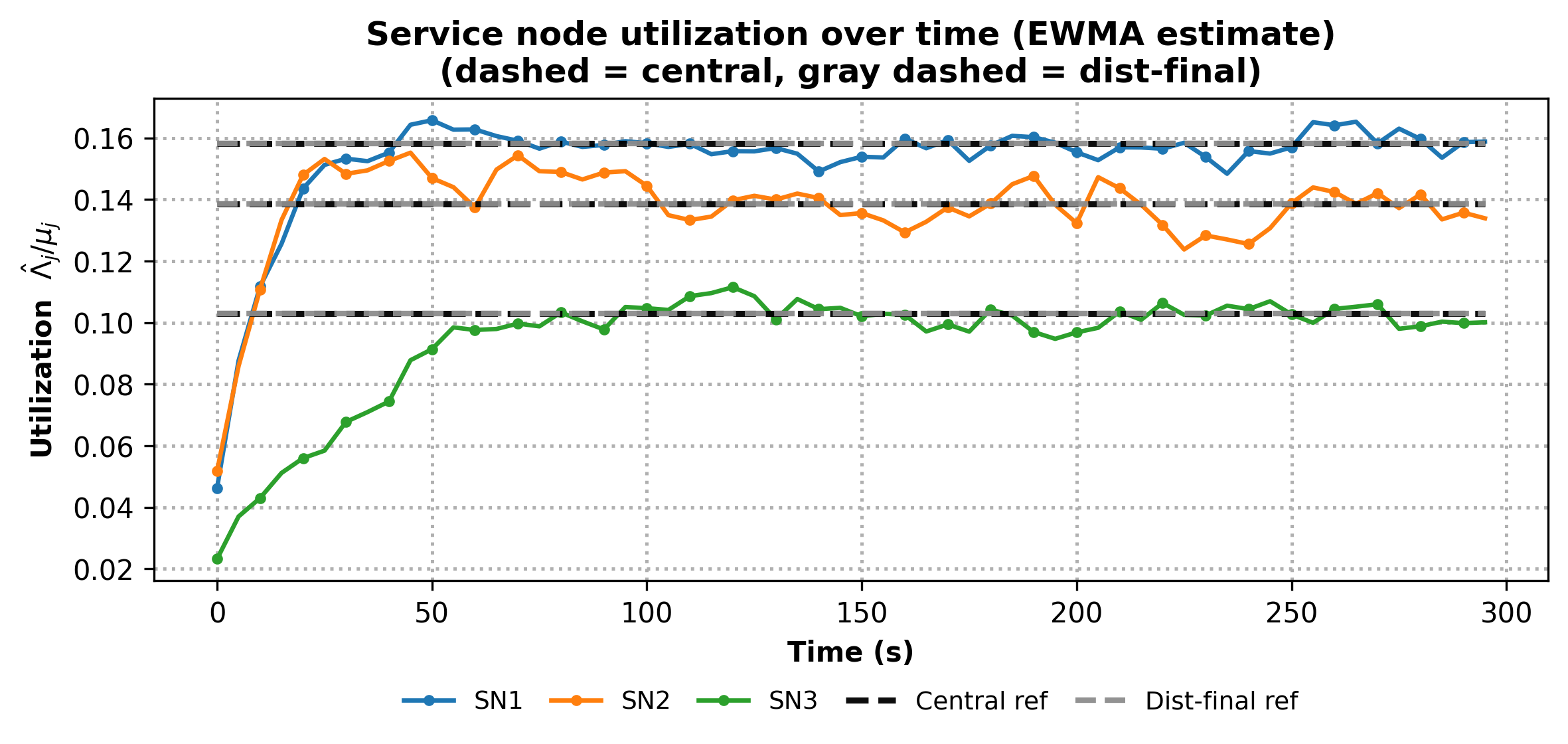}
  \caption{Windowed stochastic count experiment: EWMA offered-load utilization. Dashed lines show the nearly coincident centralized and distributed deterministic references.}
  \label{fig:stoch_brokerutil_vs_time}
\end{figure}

Figure~\ref{fig:stoch_brokerutil_vs_time} shows the EWMA offered-load utilization $\widehat\Lambda_j/\mu_j$ together with centralized and distributed deterministic references. After the warmup transient, all three trajectories fluctuate around the corresponding reference levels. Their post-warmup means are $(0.1571,0.1403,0.0984)$, giving relative differences of approximately $0.65\%$, $1.18\%$, and $4.41\%$ from the deterministic distributed utilizations for SN1--SN3. Thus the noisy windowed controller preserves the same overall load-distribution pattern over the finite run.

\setcounter{figure}{7}
\begin{figure*}[!tb]
  \centering
  \includegraphics[width=0.96\textwidth]{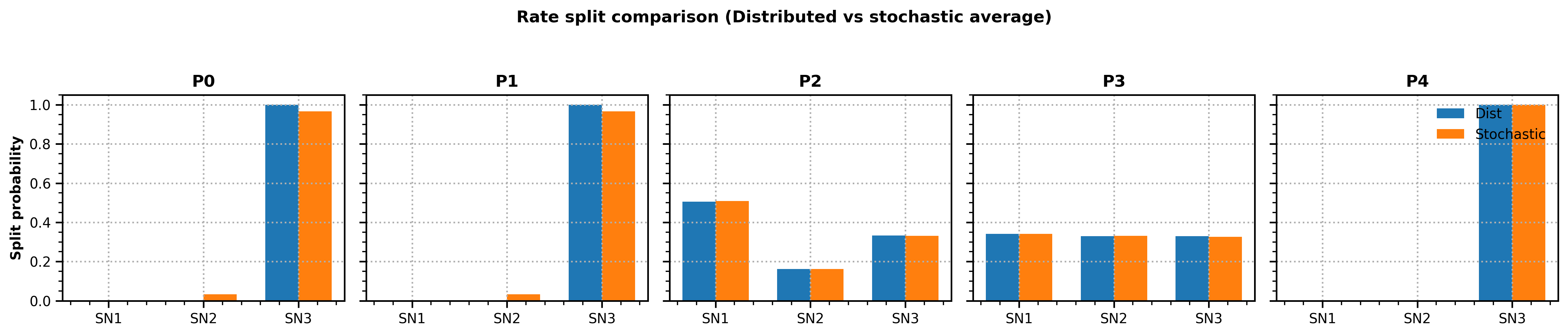}
  \caption{Deterministic fixed-point routing fractions and post-warmup window-averaged fractions for each source. The largest finite-horizon deviations occur for P0 and P1 during inertial adaptation.}
  \label{fig:ratesplit_dist_vs_stoch}
\end{figure*}

Figure~\ref{fig:stoch_persource_delay_vs_time} reports the routing-weighted queue-based proxy obtained from $(q+1)/\mu$ on each access and service stage. The curves settle after the routing transient and preserve the heterogeneous ordering seen in the deterministic experiment. In this low-utilization run the service-node queues remain zero, so the figure reflects changes in routing weights and service-rate terms rather than observed queueing sojourn times.

Figure~\ref{fig:ratesplit_dist_vs_stoch} compares the deterministic fixed-point routing fractions with the post-warmup window averages. P2 and P3, which split traffic across all service nodes, agree within $0.34$ and $0.27$ percentage points, respectively. P4 matches exactly in the stored run. P0 and P1 retain approximately $3.3\%$ on SN2 because the comparison includes a finite warmup and inertial adaptation; correspondingly, their average SN3 fractions are slightly below the deterministic value. These deviations describe finite-horizon adaptation rather than a different limiting allocation.

For precision, this experiment is a five-second windowed count process rather than an event-driven M/M/1 simulator: it does not timestamp individual jobs or schedule individual exponential service events, and byte rates are treated as normalized work rates. It therefore tests finite-window load tracking and routing adaptation under count noise. Distributional delay and tail-latency validation require an event-driven study, which we leave to future work.

\setcounter{figure}{6}
\begin{figure}[H]
  \centering
  \includegraphics[width=\linewidth]{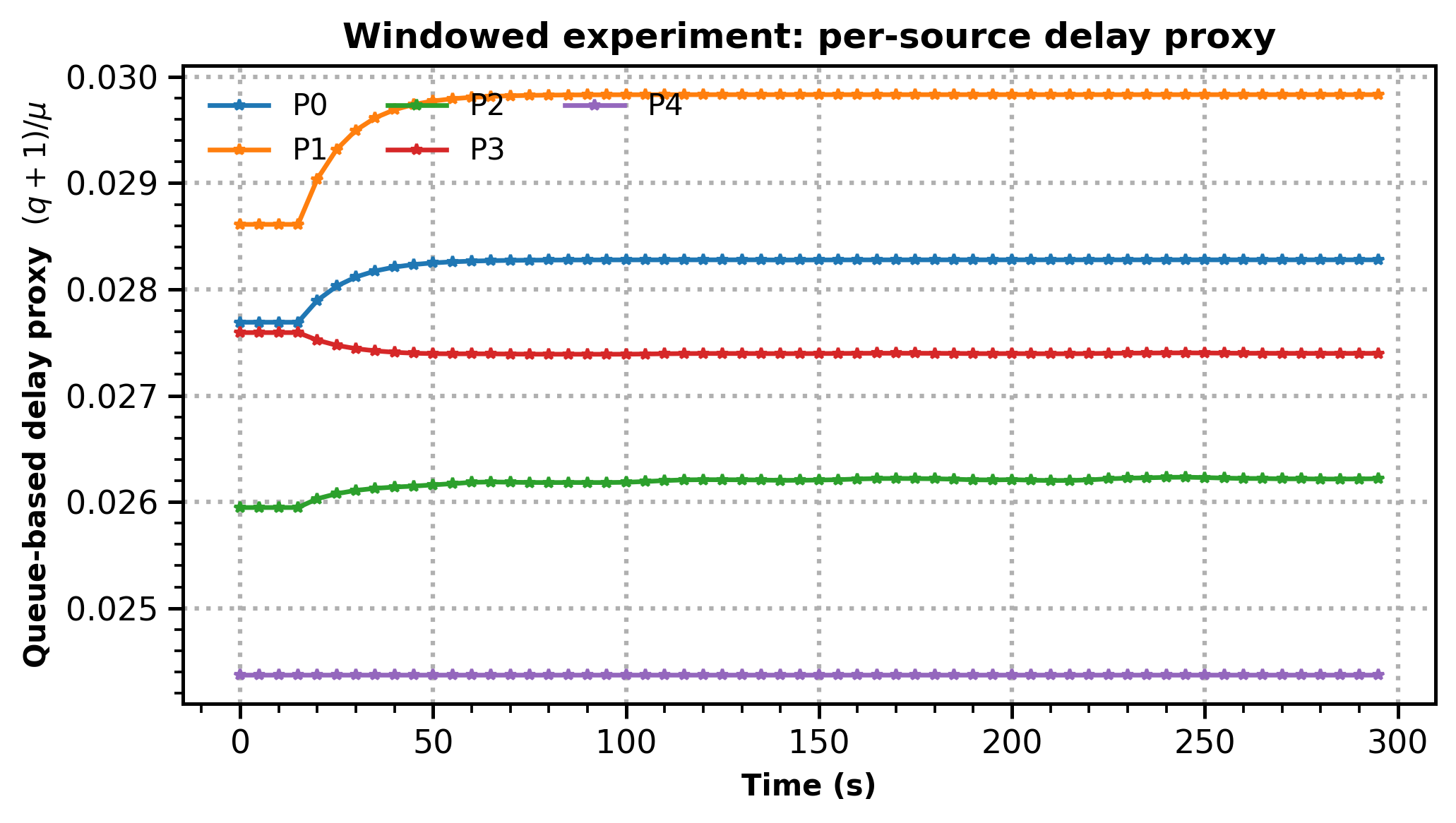}
  \caption{Windowed stochastic count experiment: per-source queue-based delay proxy versus time. The proxy uses $(q+1)/\mu$ at each stage and is not an observed packet sojourn time.}
  \label{fig:stoch_persource_delay_vs_time}
\end{figure}
\setcounter{figure}{8}

\subsection{Discussion}
\label{subsec:results_discussion}
The deterministic results establish three consistent numerical observations for the studied instance. First, the distributed objective, service-node loads, and active-route pattern converge to their centralized counterparts. Second, the independent residual battery confirms conservation, capacity feasibility, price consistency, fixed-point consistency, and KKT optimality without modifying the source response after its multiplier solve. Third, the windowed experiment shows that the same allocation structure remains visible under sampled arrival and service counts and inertial routing updates. Together these results provide a reproducible numerical realization of the fixed-point optimality characterization. They complement, but do not replace, the analytical guarantee in Proposition~1: the proposition certifies feasible fixed points, while the trajectories document the behavior of this static execution.

\section{Conclusion}
\label{sec:conclusion}
We studied a steady-state multi-source traffic allocation problem in which end-to-end delay is driven jointly by rate-dependent, capacity-constrained access paths and load-dependent queueing at capacity-constrained service nodes. Under standard monotonicity and convexity assumptions, the flow-weighted delay objective is convex and its global optimum is characterized by Wardrop-type total-marginal-cost equalization. This structure yields a lightweight distributed pricing-based algorithm in which service nodes broadcast scalar congestion prices and sources solve small separable allocation problems. A common feasibility step protects shared service-node capacities under simultaneous updates, and every feasible fixed point is globally system-optimal.

On the verified static instance, the distributed execution attains the centralized objective and service-node utilizations to numerical precision. Its conservation, capacity, price-consistency, fixed-point, and KKT residuals satisfy the prescribed tolerances. A complementary windowed stochastic count experiment also reproduces the deterministic load and routing patterns under sampled counts and inertial adaptation. These analytical and numerical results provide the synchronous foundation for studying delayed and asynchronous implementations.

Future directions include analyzing fully asynchronous implementations with delayed or stale information, establishing feasibility and delay-dependent guarantees for the selected controller, and extending the model to time-varying capacities and nonstationary operating regimes. Other directions include comparing the proposed capacity-aware optimization framework against capacity-unaware but queue-aware control policies, including Lyapunov drift-plus-penalty (backpressure) methods, and adding event-driven queueing and trace-driven evaluations. Reinforcement learning, multi-agent reinforcement learning, and hybrid model-data-driven methods that combine structural insights from convexity and marginal-cost pricing with learned components are also natural subjects for future study.

\noindent\textbf{The source code for the algorithms and numerical experiments is publicly available in the} \href{https://github.com/ANRGUSC/JointNetServerCongestion-WiOpt26}{GitHub repository}. The repository includes KKT/Wardrop checks, convergence plots, residual diagnostics, and the windowed count experiment described above.

\appendices
\section{Changes Relative to the WiOpt 2026 Version}
\label{app:corrections}
This revision preserves the central convex formulation and marginal-cost characterization of the conference paper~\cite{wiopt2026jnsc} while making the following corrections and qualifications.
\begin{enumerate}
\item \textbf{Capacity convention.} The optimization is stated consistently on the open stability domain; the numerical solver margin is distinguished from the analytical problem.
\item \textbf{Source best response.} Equation~\eqref{eq:mm1_br} starts the multiplier search at the minimum activation threshold and solves conservation directly. The erroneous post-hoc normalization is removed.
\item \textbf{Feasibility.} Algorithm~\ref{alg:pricing} adds the aggregate safe-step needed to protect coupled service capacities.
\item \textbf{Theoretical claim.} The former contraction claim is replaced by the rigorous statement that every feasible fixed point is globally optimal. Global convergence on the full open M/M/1 domain is not claimed.
\item \textbf{Numerical results.} The corrected distributed objective is $2.0157$, versus $2.0157$ centrally; figures and tables use the corrected implementation and include residual certificates.
\item \textbf{Windowed experiment.} The former ``stochastic M/M/1 validation'' is corrected to a five-second windowed stochastic count experiment for load tracking and routing adaptation; it is not described as an event-driven simulation or an observed-sojourn study.
\end{enumerate}

\section*{AI Use Statement}
The authors have used AI tools (including Claude Code and ChatGPT) for writing code and simulations, for help with design and analysis of algorithms, as well as with the paper writing, editing, proofreading, and formatting. The human authors accept full responsibility for the contents of this work.

\bibliographystyle{ieeetr}
\bibliography{references}

@article{gallager1977minimum,
  title={A minimum delay routing algorithm using distributed computation},
  author={Gallager, Robert G},
  journal={IEEE Transactions on Communications},
  volume={25},
  number={1},
  pages={73--85},
  year={1977}
}

@article{kelly1998rate,
  title={Rate control for communication networks: shadow prices, proportional fairness and stability},
  author={Kelly, Frank P and Maulloo, Aman K and Tan, David KH},
  journal={Journal of the Operational Research Society},
  volume={49},
  number={3},
  pages={237--252},
  year={1998}
}

@article{low1999optimization,
  title={Optimization flow control. I. Basic algorithm and convergence},
  author={Low, Steven H and Lapsley, David E},
  journal={IEEE/ACM Transactions on Networking},
  volume={7},
  number={6},
  pages={861--874},
  year={1999}
}

@article{rambha2018marginal,
  title={Marginal cost pricing for system optimal traffic assignment with recourse under supply-side uncertainty},
  author={Rambha, Tarun and Boyles, Stephen D and Stone, Peter},
  journal={Transportation Research Part B: Methodological},
  volume={110},
  pages={104--131},
  year={2018}
}

@article{nguyen2024network,
  title={Network-Aided Intelligent Traffic Steering in 6G O-RAN: A Multi-Layer Optimization Framework},
  author={Nguyen, Van-Dinh and others},
  journal={IEEE Transactions on Mobile Computing},
  year={2024}
}

@article{habib2023traffic,
  title={Traffic Steering for 5G Multi-RAT Deployments using Deep Reinforcement Learning},
  author={Habib, A and others},
  journal={arXiv preprint arXiv:2301.05316},
  year={2023}
}

@article{mitzenmacher2001power,
  title={The power of two choices in randomized load balancing},
  author={Mitzenmacher, Michael},
  journal={IEEE Transactions on Parallel and Distributed Systems},
  volume={12},
  number={10},
  pages={1094--1104},
  year={2001}
}

@article{lu2011jiq,
  title={Join-Idle-Queue: A novel load balancing algorithm for dynamically scalable web services},
  author={Lu, Yi and others},
  journal={Performance Evaluation},
  volume={68},
  number={11},
  pages={1056--1071},
  year={2011}
}

@article{balseiro2025load,
  title={Load Balancing with Network Latencies via Distributed Gradient Descent},
  author={Balseiro, Santiago and others},
  journal={arXiv preprint arXiv:2504.10693},
  year={2025}
}

@article{urgaonkar2015dynamic,
  title={Dynamic service migration and workload scheduling in edge-clouds},
  author={Urgaonkar, Rahul and others},
  journal={Performance Evaluation},
  volume={91},
  pages={205--228},
  year={2015}
}

@article{nguyen2021price,
  title   = {Price-based Resource Allocation for Edge Computing: A Market Equilibrium Approach},
  author  = {Nguyen, Duong Tung and Le, Long Bao and Bhargava, Vijay},
  journal = {IEEE Transactions on Cloud Computing},
  volume  = {9},
  number  = {1},
  pages   = {302--317},
  year    = {2021},
  doi     = {10.1109/TCC.2018.2844379}
}

@inproceedings{qureshi2022plb,
  title        = {PLB: Congestion Signals are Simple and Effective for Network Load Balancing},
  author       = {Qureshi, Mubashir Adnan and Cheng, Yuchung and Yin, Qianwen and Fu, Qiaobin and Kumar, Gautam and Moshref, Masoud and Yan, Junhua and Jacobson, Van and Wetherall, David J. and Kabbani, Abdul},
  booktitle    = {Proceedings of the ACM SIGCOMM 2022 Conference (SIGCOMM '22)},
  year         = {2022},
  pages        = {207--218},
  address      = {Amsterdam, Netherlands},
  publisher    = {ACM},
  doi          = {10.1145/3544216.3544226}
}

@article{paccagnan2019nash,
  title        = {Nash and Wardrop Equilibria in Aggregative Games With Coupling Constraints},
  author       = {Paccagnan, Dario and Gentile, Basilio and Parise, Francesca and Kamgarpour, Maryam and Lygeros, John},
  journal      = {IEEE Transactions on Automatic Control},
  year         = {2019},
  volume       = {64},
  number       = {4},
  pages        = {1373--1388},
  doi          = {10.1109/TAC.2018.2849946}
}

@inproceedings{vu2021fast,
  title        = {Fast Routing under Uncertainty: Adaptive Learning in Congestion Games with Exponential Weights},
  author       = {Vu, Dong Quan and Antonakopoulos, Kimon and Mertikopoulos, Panayotis},
  booktitle    = {Advances in Neural Information Processing Systems (NeurIPS)},
  year         = {2021},
  url          = {https://proceedings.neurips.cc/paper/2021/file/7b86f36d139d8581d4b5a4f155ba431c-Paper.pdf}
}

@inproceedings{wiopt2026jnsc,
  author    = {Tamoghna Sarkar and Bhaskar Krishnamachari},
  title     = {Joint Network-and-Server Congestion in Multi-Source Traffic Allocation: A Convex Formulation and Price-Based Decentralization},
  booktitle = {2026 24th International Symposium on Modeling and Optimization in Mobile, Ad Hoc, and Wireless Networks (WiOpt)},
  pages     = {1--8},
  year      = {2026},
  doi       = {10.23919/WiOpt71098.2026.11568243}
}

\end{document}